\documentclass[aps,superscriptaddress,floatfix,showpacs,amsmath,amssymb,twocolumn,pra]{revtex4-1}
\usepackage{bm}
\usepackage[dvips]{graphicx}
\usepackage{wrapfig,subfigure}
\usepackage{amssymb}
\usepackage{stmaryrd}
\usepackage{url}
\usepackage{color}
\usepackage[normalem]{ulem}
\usepackage{enumerate}
\usepackage{epstopdf}
\usepackage{comment}
\graphicspath{ {./pics/} }
\def\rb{{\bf r}}

\def\ub{{\bf u}}
\def\eb{{\bf e}}

\def\kb{{\bf k}}

\def\ep{\varepsilon}

\begin{document}
\title{Strong vibration nonlinearity in semiconductor-based nanomechanical systems}
\author{Kirill Moskovtsev}
\affiliation{Department of Physics and Astronomy, Michigan State University, East Lansing, Michigan 48824, USA}
\author{M. I. Dykman}
\affiliation{Department of Physics and Astronomy, Michigan State University, East Lansing, Michigan 48824, USA}

\date{\today}

\begin{abstract}
We study the effect of the electron-phonon coupling on vibrational eigenmodes of nano- and  micro-mechanical systems made of semiconductors with equivalent energy valleys. 
We show that the coupling can lead to a strong mode nonlinearity. The mechanism is the lifting of the valley degeneracy by the strain.  The redistribution of the electrons between the valleys is controlled by a large ratio of the electron-phonon coupling constant to the electron chemical potential or temperature. We find the  quartic in the strain terms in the electron free energy, which determine the amplitude dependence of the mode frequencies. This dependence is calculated for silicon micro-systems. It is significantly different for different modes and the crystal orientation, and can vary nonmonotonously with the electron density and temperature.
\end{abstract}

%\pacs{05.40.-a}

\maketitle

\section{Introduction}

The electron-phonon coupling strongly affects vibrational modes of nano- and micro-electro-mechanical systems. Much interest have attracted the effects of this coupling  related to the reduced dimensionality of the electron system, as they make it possible to reveal  interesting consequences of the electron correlations at the nanoscale, the Coulomb blockade being a simple example, cf.  \cite{Fedorets2004,Chtchelkatchev2004,Koch2005a,Pistolesi2007,Usmani2007,Leturcq2009,Steele2009,Lassagne2009,Kiselev2013} and references therein. 

Much less attention has been paid to the consequences of the electron-phonon coupling, which are related to the discreteness of the vibrational spectrum of a nanosystem, but emerge in the absence of size quantization of the electron motion. One of such consequences, which we study in this paper, is the coupling-induced change of the vibration nonlinearity.  Strong nonlinearity is a generic feature of vibrations in small systems \cite{Cleland2003,Lifshitz2008}. Its easily accessible  manifestation is the dependence of the mode frequencies on the vibration amplitudes. This dependence corresponds to the ``self-action" of the mode, and its familiar analog in bulk crystals are acoustic solitons \cite{Hao2001,Akimov2007}; however, the nonlinearity required for observing such solitons usually is sufficiently strong only for high-frequency phonons. Also, the change of the eigenfrequency with the mode amplitude is of interest for modes with a discrete frequency spectrum, such as standing waves in mesoscopic systems, but not for propagating waves with a quasi-continuous spectrum.

Much attention have been recently attracting Si-based nano- and micromechanical systems, see \cite{Ghaffari2013,Sansa2016} and references therein. In such systems there was observed an unexpectedly large change of the amplitude dependence of the vibration frequency  with the varying electron density \cite{shahmohammadi2013nonlinearity,Yang2016}.  When the doping level was increased from $2.8\times 10^{18}$~cm$^{-3}$ to   $5.9\times 10^{19}$~cm$^{-3}$, the nonlinearity parameter increased by more than an order of magnitude. Moreover, the nonlinearity change was different for the vibrational modes with different spatial structure. 

In this paper we develop a theory of the nonlinearity of vibrational modes in semiconductor nano- and micro-mechanical systems with high electron density. We show that the electron-phonon coupling can lead to a strong self-action of the vibrational modes, which in turn significantly modifies the amplitude dependence of the mode frequencies. We find the dependence of the effect on the electron density and temperature.

For bulk semiconductors, the effect of the electron-phonon coupling on the elastic properties, including the three-phonon coupling, was first analyzed by Keyes \cite{keyes1961electronic}. The analysis referred to $n$-Ge and was based on the deformation potential approximation. The idea was that deformation lifts the degeneracy of the equivalent electron valleys, which leads to a redistribution of the electrons over the valleys. In turn, such redistribution changes the speed of sound depending on the direction and polarization of the sound waves and also affects the sound speed in the presence of uniaxial stress. This theory was extended to silicon and the corresponding measurements were done by Hall \cite{Hall1967}. However, Hall also observed the change of the speed of transverse sound waves and the effect of stress on sound propagation in the geometries, where these effects are due to shear deformation and do not arise in the deformation potential model.  A theory of the change of the linear shear elastic constant in silicon due to the intervalley redistribution of the electrons was developed by Cerdeira and Cardona \cite{cerdeira1972effect}.

As we show, in mesoscopic systems the strain-induced redistribution of the electrons over the valleys of the conduction band leads to the previously unexplored strong fourth-order nonlinearity of the vibrational modes. This nonlinearity gives a major contribution to the amplitude dependence of the vibration frequency. The redistribution also leads to a temperature dependence of the frequencies. The magnitudes of the effects sensitively depends on the mode structure. We describe them for several types of modes, including those studied in the experiment \cite{shahmohammadi2013nonlinearity,Yang2016} and qualitatively compare the results with the observations. The theoretical results refer to both degenerate and nondegenerate electron systems. Specific calculations are done for silicon resonators.  

In Sec.~\ref{sec:model} we give, for completeness, the expressions for the mode normalization and the amplitude-dependent frequency shift of coupled nonlinear modes in a nano- or micro-system. In Sec.~\ref{sec:e-ph_coupling} and Appendix~\ref{AppendixA} we provide expressions for the electron-phonon coupling induced change of the elasticity parameters, including the parameters of quartic nonlinearity. In Sec.~\ref{sec:explicit_form} we discuss the asymptotic behavior of the parameters of quartic nonlinearity for low and high electron density and give their explicit form for silicon. In Sec.~\ref{sec:simple_modes} we calculate the nonlinear frequency shift for several frequently used vibrational modes in single-crystal silicon systems  and show the dependence of this shift on the electron density and temperature. The explicit analytical expressions are given in Appendices~\ref{sec:Lame} and \ref{sec:extension}. Sec.~\ref{sec:conclusions} contains concluding remarks. 

%%%%%%%%%%%%%%%%%%%%%%%%%%%%%
%%%%%%%%%%%%%%%%%%%%%%%%%%%%%

\section{Nonlinear frequency shift of low-frequency eigenmodes}
\label{sec:model}

Of primary interest for nano- and micro-mechanical systems are comparatively low-frequency modes with wavelength on the order of the maximal size of the system. Examples are provided by long-wavelength flexural modes of nanotubes, nanobeams, and nano/micro-membranes, or acoustic-type modes in microplates or beams. These modes are easy to excite and detect. We will enumerate them by index $\nu$. Their dynamics is described by the elasticity theory \cite{LL_Elasticity}. The spatial structure of the displacement field of a mode $\ub^{(\nu)}(\rb)$ in the harmonic approximation is determined by the boundary conditions. We will choose $\ub^{(\nu)}(\rb)$ dimensionless, so that in our finite-size system 
\begin{align}
\label{eq:normalization}
\int d \rb \, \ub^{(\nu)}(\rb)\cdot \ub^{(\nu')}(\rb) =V \delta_{\nu \nu'}.  
\end{align}
Here, $V$ is the volume of the system. We assumed that the mode eigenfrequencies $\omega_\nu$ are nondegenerate; including degenerate modes is straightforward. For simplicity, we also assumed that the system is spatially uniform; an extension to spatially nonuniform systems is straightforward as well. 

We emphasize the distinction of the normalization (\ref{eq:normalization}) from the conventional normalization for bulk crystals, where $\nu$ corresponds to the wave vector and the branch number, and the normalization integral is independent of the volume. The normalization (\ref{eq:normalization}) is convenient for the analysis of low-frequency modes with the discrete spectrum characteristic of  mesoscopic systems. Such modes are standing waves, and therefore vectors $\ub^{(\nu)}$ can be chosen real.

The low-frequency part of the displacement can be written as
\begin{align}
\label{eq:displacement}
\ub (\rb,t) = \sum_\nu Q_\nu(t)\ub^{(\nu)}(\rb).
\end{align}
Functions $Q_\nu(t)$ give the mode amplitudes.  In the harmonic approximation the dynamics of the standing waves is described by the Hamiltonian
\begin{align}
\label{eq:harmonic_Hamiltonian}
&H_h = \frac{1}{2}\sum_\nu (M^{-1}P_\nu^2 +M \omega_\nu^2 Q_\nu^2),
\end{align}
where $P_\nu$ is the momentum of mode $\nu$ and  $M$ is  the mass of the system. 

The anharmonicity of the crystal leads to mode-mode coupling. Within the elasticity theory this coupling is described by the terms in the Hamiltonian, which are cubic and quartic in the strain tensor. We will not consider higher-order terms, which are small for the mode amplitudes of interest. From the expansion \eqref{eq:displacement}, we obtain the nonlinear part of the Hamiltonian in the form
\begin{align}
\label{eq:anharmonic_Hamiltonian}
H_{\rm nl} = &\frac{1}{3}\sum \beta_{\nu_1\nu_2\nu_3}Q_{\nu_1}Q_{\nu_2}Q_{\nu_3}\nonumber\\
 &+ \frac{1}{4}\sum \gamma_{\nu_1\nu_2\nu_3\nu_4}Q_{\nu_1}Q_{\nu_2}Q_{\nu_3}Q_{\nu_4}.
\end{align}
Equation \eqref{eq:anharmonic_Hamiltonian} is essentially an expansion in the ratio of the mode amplitudes to  their characteristic wavelength, which is of the order of the appropriate linear dimension of the system. This is why mesoscopic systems are of particular interest, as here vibrations of low-frequency eigenmodes become nonlinear for already small vibration amplitudes.

A familiar consequence of nonlinearity in nano- and micromechanical systems is the dependence of the vibration frequency of a mode on its own amplitude and on the amplitudes of other modes, see Ref.~\onlinecite{Lifshitz2008} for a review. In particular, the  change $\delta\omega_\nu$ of the mode frequency due to the vibrations of the mode itself, $Q_\nu(t) = A_\nu\cos \omega_\nu t$,  is \cite{LL_Mechanics2004,Dykman1973}
\begin{align}
\label{eq:Duffing_shift}
\delta\omega_\nu \approx \left[\frac{3\gamma_\nu}{8M\omega_\nu} - \sum_{\nu'}\frac{\beta_{\nu\nu\nu'}^2(3\omega_{\nu'}^2 - 8\omega_{\nu}^2)}{4M^2\omega_{\nu}^3(\omega_{\nu'}^2 - 4\omega_{\nu}^2)}\right]A_\nu^2,
\end{align}
where $\gamma_\nu \equiv \gamma_{\nu\nu\nu\nu}$ and we kept the terms of the first order in $\gamma$ and the second order in $\beta$. 

The nonlinear mode coupling (\ref{eq:anharmonic_Hamiltonian}) leads also to the frequency shift due to thermal vibrations of the modes. The dominating contribution to this shift for low-frequency modes comes from their coupling to modes with frequencies $\sim k_BT/\hbar$, which have a much higher density of states. This shift is described by an expression that is similar to Eq.~(\ref{eq:Duffing_shift}) with $A_\nu^2$  replaced by $A_{\nu'}^2 \sim k_BT/M\omega_{\nu'}^2$ and placed under the sum over $\nu'$, in the classical limit.

%%%%%%%%%%%%%%%%%%%%%%%%%%%%%%%%%%%%%
%%%%%%%%%%%%%%%%%%%%%%%%%%%%%%%%%%%%%

\section{The nonlinearity due to the electron-phonon coupling}
\label{sec:e-ph_coupling}

We will consider the vibration nonlinearity due to the electron-phonon coupling in multi-valley semiconductors with cubic symmetry, silicon and germanium being the best known examples. In such semiconductors, the energy valleys of the conduction band are located at high-symmetry axes of the Brillouin zone. Strain lifts the symmetry and thus the degeneracy of the valleys. 

The simplest mechanism of the electron-phonon coupling is the deformation potential. Here, the energy shift $\delta E_\alpha$ of valley $\alpha$ is determined by the deformation potential parameters $\Xi_u$  and $\Xi_d$ of the coupling to a uniaxial strain along the  symmetry axis of the valley and to dilatation, respectively. In terms of the strain tensor $\ep_{ij}$ we have $\delta E_\alpha = \sum_{ij}\Xi^{(\alpha)}_{ij}\ep_{ji}$, where  $\hat \Xi^{(\alpha)} = \Xi_d\hat I + \Xi_u \eb^{(\alpha)}\otimes \eb^{(\alpha)}$, with $\eb^{(\alpha)}$ being the unit vector along the symmetry axis of the valley. We use the hat symbol to indicate tensors and symbol ``$\otimes$" to indicate tensor products. The analysis below is not limited to the deformation potential approximation. An important extension will be discussed using silicon as an example.

We assume that the strain varies in time and space slowly compared to the reciprocal rate of intervalley electron scattering and the intervalley scattering length, respectively. Then the electron system follows the strain adiabatically. The electron density $n^{(\alpha)}(\rb)$ in valley $\alpha$ is decreased or increased depending on whether the bottom of the valley goes up or down. In the single-electron approximation and for the deformation potential coupling, the electron free energy density for a given strain is $F_e = \sum_\alpha \{f_e[n^{(\alpha)}(\rb)] + n^{(\alpha )}(\rb) \Xi^{(\alpha)}_{ij} \ep_{ji}(\rb)\}$ where $f_e[n(\rb)]$ is the free energy density for electrons with density $n(\rb)$ in a valley in the absence of coupling to phonons. 

The electro-neutrality requires that the total electron density summed over the valleys be constant. The free energy density $F_e$ has to be minimized over $n^{(\alpha)}(\rb)$ to meet this constraint. This gives the change of the electron chemical potential $\delta\mu$ due to strain $\hat\ep$. The resulting increment of the electron free energy density has the form of a series expansion in the strain tensor, 
\begin{align}
\label{eq:free_energy_expansion}
\delta F_e =& \widehat \Lambda_1 \cdot\hat \ep+\frac{1}{2} \widehat \Lambda_2 \cdot\hat \ep\otimes \hat\ep +\frac{1}{6} \widehat\Lambda_3  \cdot \hat \ep\otimes \hat\ep \otimes\hat \ep\nonumber\\
& +\frac{1}{24}\widehat\Lambda_4 \cdot \hat \ep\otimes \hat\ep\otimes \hat \ep\otimes \hat\ep +...
\end{align} 
Here $\widehat\Lambda_1,\widehat \Lambda_2, \widehat \Lambda_3$, and $\widehat \Lambda_4$ are tensors of ranks 2, 4,  6, and 8, respectively. They are contracted with the tensor products of the strain tensor $\hat\ep$. Respectively, $\widehat\Lambda_k$ are the electronic contributions to the linear (for $k=2$) and nonlinear (for $k>2$) elasticity parameters of the crystal. These contributions are isothermal, but since the change of the mode frequencies from the electron-phonon coupling is small and the nonlinearity is also small, the difference with the adiabatic expressions can be disregarded.

To the third order in $\hat\ep$ the expression for $\delta F_e$  in terms of the shift of the valleys was found by Keyes \cite{keyes1961electronic} in the analysis of sound wave propagation. However, to find the parameters of the quartic nonlinearity of resonant modes in small systems, which is of primary interest to us, we also need to keep quartic terms in Eq.~(\ref{eq:free_energy_expansion}).

As seen from the explicit form of the parameters of the expansion (\ref{eq:free_energy_expansion}) given in Appendix \ref{AppendixA}, $\widehat \Lambda_k \propto \Xi_u [\Xi_u/\max (\mu_0, k_BT)]^{k-1}$ ($k=1,2,...$), where $\mu_0$ is the electron chemical potential in the absence of strain;  it is determined by the total (summed over the valleys) electron density $n$.  Of central importance for the analysis is that parameter $\Xi_u/\max (\mu_0, k_BT) \sim 10^3$ for electron densities $n\sim 10^{19}~{\rm cm}^{-3}$ and room temperatures, i.e.
\begin{align}
\label{eq:strong_coupling}
\Xi_u/\max (\mu_0, k_BT) \gg 1.
\end{align}
As a consequence, the coefficients at the nonlinear in $\hat\ep$ terms in Eq.~(\ref{eq:free_energy_expansion}) quickly increase with the increasing order of the nonlinearity [the overall series (\ref{eq:free_energy_expansion}) is converging fast because of the smallness of the strain tensor]. 

The increase of  $\widehat\Lambda_k$ with $k$ allows us to
keep in $\hat\ep$ only the terms linear in the lattice displacement, i.e., to set $\ep_{ij}=(1/2)(\partial u_i/\partial x_j + \partial u_j/\partial x_i)$, where $u_i$ and $x_i$ are the components of the displacement and the coordinates, respectively. Indeed, in this case a $k$th term of the series (\ref{eq:free_energy_expansion}) is of order $k$ in the displacement.  If we included the quadratic in  $\partial u_i/\partial x_j$ term into one of the $\hat\ep$ tensors in the $k$th term, this term would become of order $k+1$ in the displacement. However, for linear $\hat\ep$ the $(k+1)$th term in the series  (\ref{eq:free_energy_expansion}) is also of the $(k+1)$th order in the displacement, but is larger by factor $\Xi_u/\max (\mu_0, k_BT)$.  

For linear $\hat\ep$, the total strain is a sum of partial contributions of strain from individual modes.  For mode $\nu$, such partial contribution is expressed in terms of the scaled displacement $\ub^{(\nu)}(\rb)$ [see Eq.~(\ref{eq:displacement})] as $\hat\ep=Q_\nu\hat\ep^{(\nu)}$, where $\ep_{ij}^{(\nu)}(\rb) = \tfrac{1}{2}[\partial u^{(\nu)}_i(\rb)/\partial x_j + \partial u^{(\nu)}_j(\rb)/\partial x_i]$. We note that, in contrast to the dimensionless strain tensor $\hat\ep$, tensor $\hat\ep^{(\nu)}$ has dimension [length]$^{-1}$. 

From Eq.~(\ref{eq:free_energy_expansion}) we find the electronic contributions to the nonlinearity parameters $\beta^{\rm (e)}_{\nu_1\nu_2\nu_3}, \gamma^{\rm(e)}_{\nu_1\nu_2\nu_3\nu_4}$  in Hamiltonian (\ref{eq:anharmonic_Hamiltonian}) , 
\begin{align}
\label{eq:nonlin_params}
\beta_{\nu_1\nu_2\nu_3}^{\rm(e)} =&\frac{1}{2}\int d\rb\, \widehat \Lambda_3 \cdot \hat\ep^{(\nu_1)}\otimes \hat\ep^{(\nu_2)}\otimes \hat\ep^{(\nu_3)}, \nonumber \\
\gamma_{\nu_1\nu_2\nu_3\nu_4}^{\rm (e)} =
&\frac{1}{6}\int d\rb\, \widehat \Lambda_4 \cdot \hat\ep^{(\nu_1)}\otimes \hat\ep^{(\nu_2)}\otimes \hat\ep^{(\nu_3)}\otimes \hat\ep^{(\nu_4)},
\end{align}
where $\hat\ep^{(\nu)}\equiv \hat\ep^{(\nu)}(\rb)$; tensors $\widehat \Lambda_k$ are independent of $\rb$. 

Similarly, the electronic contribution to the eigenfrequency is
\begin{align}
\label{eq:frequency_change}	
\Delta \omega_{\nu}^{\rm(e)} = \frac{1}{2M\omega_\nu}\int d\rb\, \widehat \Lambda_2 \cdot \hat\ep^{(\nu)}\otimes \hat\ep^{(\nu)}.
\end{align}
Generally, the term $\propto\widehat\Lambda_2$ leads to mode mixing; however, if the mode frequencies are nondegenerate, this mixing is weak and can be disregarded, to the leading order in the electron-phonon coupling. One can see that the effect of the static stress $\propto \widehat \Lambda_1$ can be disregarded as well.

The frequency change (\ref{eq:frequency_change}) depends on temperature because of the temperature dependence of $\widehat\Lambda_2$. The nonlinearity (\ref{eq:nonlin_params}) also leads to a temperature dependence of the mode eigenfrequency. Together they modify the temperature dependence of the mode eigenfrequencies compared to that of undoped crystals. This modification often weakens the temperature dependence of the eigenfrequencies, which proves very important for applications of micro-mechanical systems in devices that work in a broad temperature range \cite{Ng2015}.

Equations (\ref{eq:free_energy_expansion}) - (\ref{eq:frequency_change}) are generic and apply beyond the deformation potential approximation. This is of particular importance for silicon. Here, the electron band valleys lie on the $\langle 100\rangle$-axes close to the $X$-points on the zone boundaries where two electron energy bands cross. Lattice strain can lead to a  band splitting at $X$-points and a shift of the valleys \cite{Bir1974,Hensel1965}. Importantly, this shift results from a shear strain, which does not lead to a linear in the strain shift in the deformation potential approximation. The valley shift is quadratic in $\hat\ep$ in this case, as explained in Appendix~\ref{AppendixA}, which corresponds to an effectively two-phonon coupling. The coupling parameter $\Xi_{\rm sh}$ is quadratic in the strain-induced band splitting, see Eq.~(\ref{eq:general_shift}). It is large, much larger than the constant $\Xi_u$. Therefore the arguments given below Eq.~(\ref{eq:strong_coupling}) apply in this case as well. For purely shear strain in silicon, terms of odd order in $\hat\ep$ in $\delta F_e$, Eq.~(\ref{eq:free_energy_expansion}),  vanish.

%%%%%%%%%%%%%%%%%%%%%%%%%%%%%%%%%%%%%%%

\begin{center}
	\begin{table*}[t]
		\caption{The change of the components of the nonlinear elasticity tensors due to the strain-induced electron redistribution between equivalent energy  valleys in doped silicon. The coordinate axes are chosen along the $\langle 100\rangle$ axes. Parameter $\Xi_{\rm sh}$ characterizes the effectively two-phonon coupling to shear strain. This parameter as well as function $F_{1/2}(x)$ are defined in Appendix~\ref{AppendixA}; $x=\mu_0/k_BT$ and $n$ is the electron density. }		
		\begin{tabular}{ccccl}
			\hline\hline
			$\delta c_{144} = -2\delta c_{155}$ &\hspace*{8pt}& $\frac{1}{9}n \Xi_{u}\Xi_{\rm sh} C_1$&\hspace*{8pt}& $C_1= F_{1/2}'/ F_{1/2} k_B T=d\ln n/d\mu_0$\\[8pt] 
			$\delta c_{1111} = - 2\delta c_{1112} 
			=2\delta c_{1122}$
			& \hspace*{8pt}& $\frac{2}{27} n \Xi_{u}^{4} C_2$&\hspace*{8pt}& $C_2= (k_BT)^{-3}F_{1/2}^{\prime \,2} \left[d^2(1/F_{1/2}')/dx^2\right]/F_{1/2}$\\[8pt]
&&&& \hspace*{15pt}$=
(dn/d\mu_0)^2 \left[d^2(d\mu_0/dn)/d\mu_0^2\right]/n$\\ [8pt]
						$\delta c_{1144}  = -2\delta c_{1155} =-2\delta c_{1244} =  \delta c_{1266}$
			&\hspace*{18pt}& $-\frac{1}{27}n \Xi_{u}^{2}\Xi_{\rm sh}C_3$&\hspace*{8pt}& $C_3= F_{1/2}''/   F_{1/2} (k_B T)^2 =n^{-1}d^2n/d\mu_0^2$ \\ [8pt]
			$\delta c_{4444}  = -6\delta c_{4455}$&\hspace*{8pt}& $-  \frac{1}{6}n \Xi_{\rm sh}^2C_4$&\hspace*{8pt}& $C_4= F_{1/2}'/F_{1/2} k_B T= d \ln n/d\mu_0 $\\ 
			\hline\hline
		\end{tabular}
		\label{table:tensors}
	\end{table*}
\end{center}

%%%%%%%%%%%%%%%%%%%%%%%%%%
%%%%%%%%%%%%%%%%%%%%%%%%%

\section{Explicit form of the tensors of nonlinear elasticity}
\label{sec:explicit_form}
Tensors $\widehat\Lambda_n$ can be obtained by minimizing the free energy density of the electron system for a given strain and expanding the result in a series in $\hat\ep$. A general procedure that allows one to find the components $\widehat\Lambda_n$ for $n\leq 4$ is described in Appendix~\ref{AppendixA}.  Using the symmetry arguments, the elasticity tensors are conveniently written in the contracted (Voigt) notation where the symmetric strain tensor is associated with a six-component vector. Then the nonlinear elasticity tensors  $\widehat \Lambda_3$ and  $\widehat\Lambda_4$ become tensors of rank three and four in the corresponding vector space. We use notation $\delta \hat c $ for tensors $\widehat\Lambda$ in these notations to emphasize that we are calculating  corrections to the nonlinear elasticity tensors due to the electron-phonon coupling.

The explicit expressions for the nonlinear elasticity tensors $\delta\hat c$ are given in Table~\ref{table:tensors}. They refer to silicon and include the contributions that come from both the deformation potential coupling and from the splitting of the electron bands due to shear strain. In the deformation potential approximation, the components of the third-rank tensor $\delta\hat c$, which determine the cubic in the strain terms in the free energy, were found earlier \cite{Hall1967}. Therefore we give only the components that contain a contribution from shear strain. 

The fourth-rank tensor $\delta\hat c$ determines the quartic in the strain terms in the free energy and has not been discussed before, to the best of our knowledge. We give all independent components of this tensor. It is expressed in terms of the derivative of the electron density $n$ over the chemical potential in the absence of strain $\mu_0$, which is a familiar thermodynamic characteristic. It is intuitively clear that the considered effect of the change of the electron density in different valleys in response to strain should be related to the derivative $dn/d\mu_0$. Interestingly, because we consider nonlinear response to strain, the expressions in Table~\ref{table:tensors} contain also higher-order derivatives of $n$ over $\mu_0$. As we will see, this leads to a nontrivial behavior of the nonlinear frequency shift with varying temperature and density. The considered mechanism of the strain-induced inter-valley electron redistribution does not contribute to the components $c_{1123}$ and $c_{1456}$, therefore $\delta c_{1123} = \delta c_{1456}=0$.

%%%%%%%%%%%%%%%%%%%%%%%%%%%%%%%%%%%%%%%%%%

\subsection{Nonlinear elasticity in the limiting cases}
\label{sec:limiting}

The expressions for $\delta\hat c$ simplify in the case of low doping (or high temperature), where the electron gas is strongly nondegenerate, and in the opposite case of a strongly degenerate electron gas. For a nondegenerate gas, where the chemical potential in the absence of strain is $\mu_0 <0, |\mu_0|\gg k_BT$, we have in Table~\ref{table:tensors} $F_{1/2}(x)= \tfrac{1}{4}\pi^{1/2}e^{x}$ with $x=\mu_0/k_BT$. The $\mu_0$-dependent factors $\exp(\mu_0/k_BT)$ in $F_{1/2}$ and its derivatives cancel each other in the expressions for $\delta\hat c$  and drop out from these expressions. The dependence of $\delta\hat c$ on density is then just linear, $\delta\hat c\propto n$. Parameters $C_{1,..., 4}$ in Table~\ref{table:tensors} depend only on temperature, $C_1\propto T^{-1}, C_2 \propto T^{-3}, C_3\propto T^{-2}$ and $C_4 \propto T^{-1}$.

The decrease of the nonlinear elasticity parameters with increasing temperature in a nondegenerate electron gas is easy to understand. The effect we consider is determined by the competition between the energetically favorable unequal population of the electron energy valleys in a strained crystal and the entropically more favorable equal valley population. With increasing temperature the entropic factor becomes stronger, leading to a smaller population difference and thus smaller effect of the electron system on the vibrations. 

For strong doping, where $\mu_0/k_BT\gg 1$, we have $\mu_0\propto n^{2/3}$, and then $F_{1/2}(x)\approx \tfrac{2}{3}x^{3/2}$ with $x=\mu_0/k_BT$. Therefore parameters $C_{1,...,4}$ in Table~\ref{table:tensors} become temperature independent, with $nC_1\propto n^{1/3}, nC_2\propto n^{-1}, nC_3\propto n^{-1/3}$, and $C_4\propto n^{1/3}$. 

The results on the asymptotic behavior of the corrections to nonlinear elasticity are not limited to silicon. Since parameters $C_{1,2,3,4}$ are given by the coefficients in the general expansion of the free energy in strain, (\ref{eq:total_F}), these results can be applied to the nonlinear elasticity induced by the  electron-phonon coupling  in other multi-valley semiconductors. To illustrate this point, in Appendix~\ref{sec:germanium} we give  $\delta\hat c$ tensor in germanium.

The difference between the asymptotic behavior of the tensors $\delta\hat c$ in the limits of nondegenerate and strongly degenerate electron gas can lead to a peculiar density and temperature dependence of the nonlinear frequency shift of the vibrational modes. It comes from the coefficients $C_{1,...,4}$ containing higher-order derivatives of $n$ with respect to $\mu_0$. In the transition region $\mu_0\sim k_BT$, thinking  of the competition between the entropic and energetic factors does not provide a simple insight into the behavior of $\delta\hat c$, as both the energy and the entropy are complicated functions of density and temperature.

%%%%%%%%%%%%%%%%%%%%%%%%%%%%%%%%%%%%%%%%%%%%%%%%
%%%%%%%%%%%%%%%%%%%%%%%%%%%%%%%%%%%%%%%%%%%%%%%%

\section{Doping-induced nonlinearity of simple vibrational modes}
\label{sec:simple_modes}
The nonlinear elasticity tensors in Table~\ref{table:tensors} give the doping-induced contributions to the nonlinearity parameters of the eigenmodes of micro- and nanomechanical systems. These contributions are described by Eq.~(\ref{eq:nonlin_params}). As mentioned before, an important characteristic of the mode nonlinearity is the dependence of the mode frequency on the vibration amplitude. To the leading order, it is given by Eq.~(\ref{eq:Duffing_shift}). This dependence has a contribution from the nonlinearity of an undoped crystal, which is quadratic in the parameters of the cubic nonlinearity; for example, if the latter is described by the Gr\"uneisen constant, the corresponding contribution is quadratic in this constant. It is typically small. There is also a contribution from the quartic nonlinearity; the parameters of such nonlinearity are not known in undoped crystals and are not expected to be large. Respectively, the amplitude dependence of the vibration frequency for low-frequency modes in weakly doped single-crystal micro-mechanical systems is relatively weak \cite{Yang2016}.

\begin{figure*}[t]
	\includegraphics[width = 0.4\textwidth]{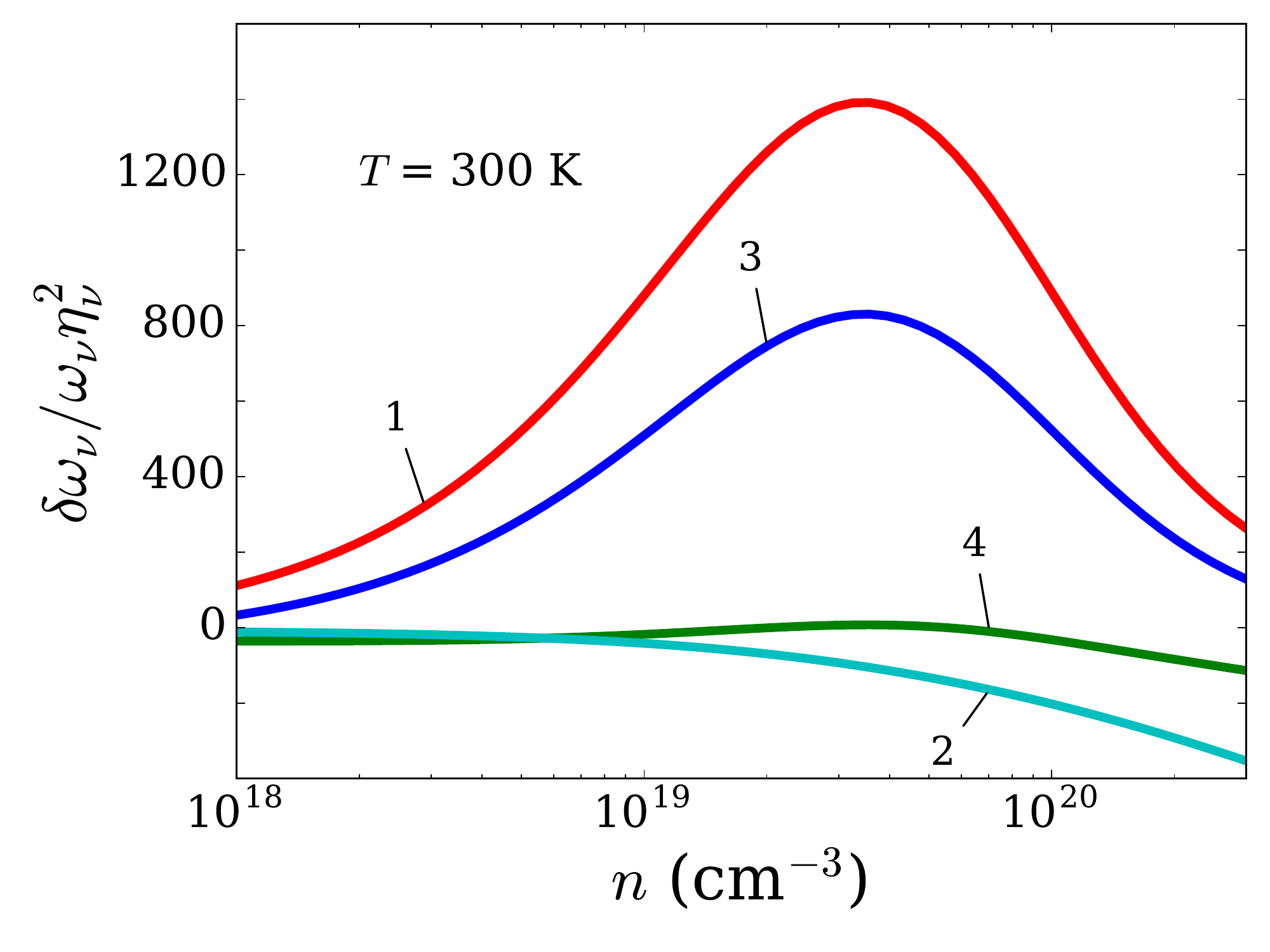}\hfill
\includegraphics[width = 0.4\textwidth]{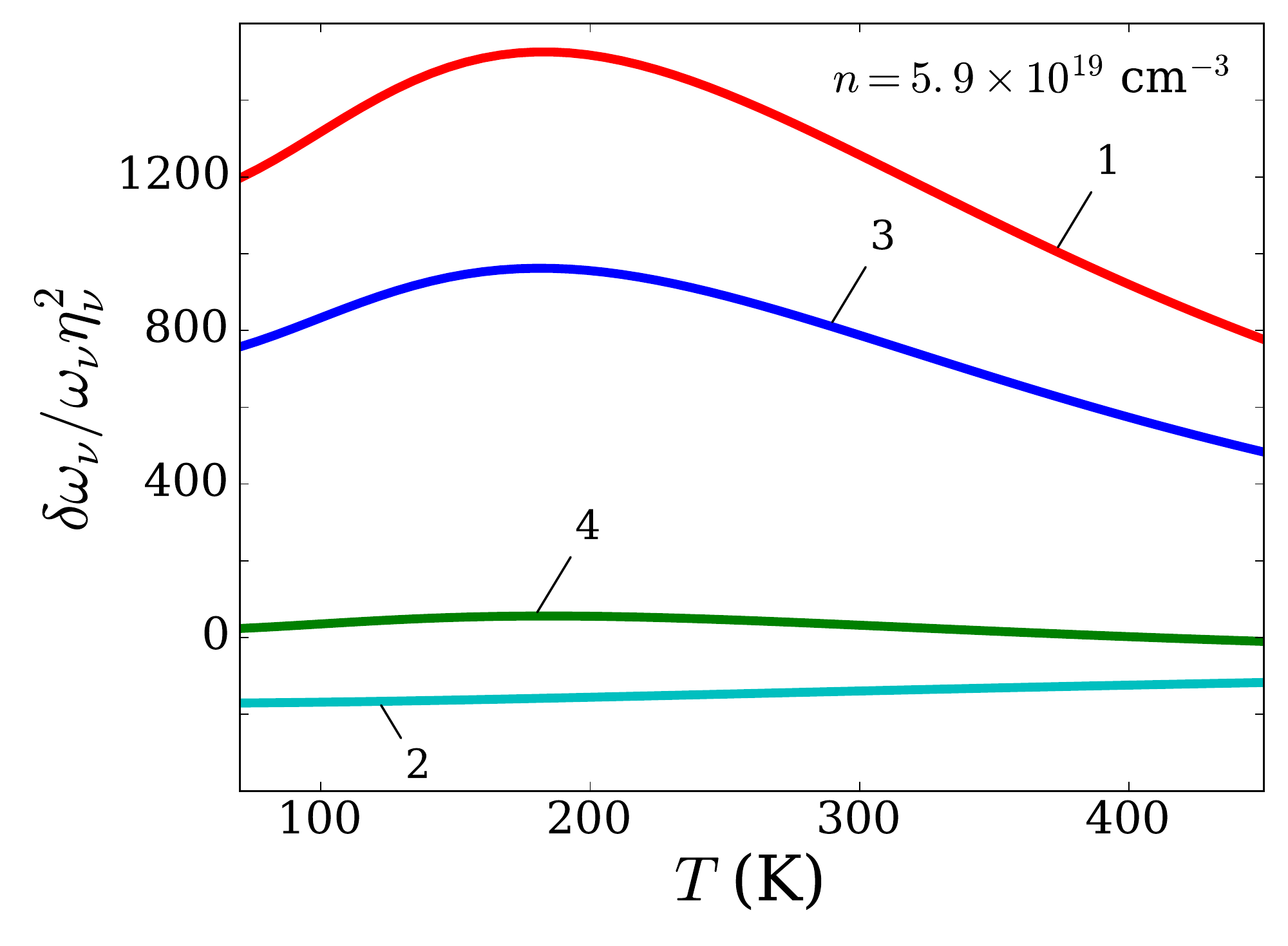}
\caption{Relative change $\delta\omega_\nu/\omega_\nu$ of the vibration frequency of a mode with the vibration amplitude  $\eta_\nu$ scaled by the relevant size of the system, cf. Eq.~(\ref{eq:scaling}). The results refer to single crystal silicon resonators. Curves 1 and 2 refer to the first Lam\'e mode in square plates cut in $\langle100\rangle$ and $\langle 110\rangle$ directions, respectively. In this case, the size of the resonator  is the length of the side of the square. Curves 3 and 4 refer to the first extension mode in beams cut in  $\langle100\rangle$ and $\langle 110\rangle$ directions, respectively. In this case, the size of the resonator  is the length of the beam.}
\label{fig:n_dependence}
\end{figure*}

A feature of the doping-induced nonlinearity described by Table~\ref{table:tensors} is that the quartic in the strain term in the free energy has a large coefficient compared to the cubic term, cf. Eq.~(\ref{eq:strong_coupling}) and the discussion below this equation. 
%
%Therefore, the  amplitude dependence of the vibration frequency  due to doping is given by Eq.~(\ref{eq:Duffing_shift}) in which one can keep only the Duffing nonlinearity constant $\gamma_\nu$, whereas the contribution from the cubic nonlinearity terms $\propto \beta_{\nu\nu\nu'}^2$ can be disregarded.
Therefore, in Eq.~(\ref{eq:Duffing_shift}) for the amplitude dependence of vibration frequency one can keep only the Duffing nonlinearity constant $\gamma_\nu$. The contribution from the cubic nonlinearity terms $\propto \beta_{\nu\nu\nu'}^2$ can be disregarded.
For a mode $\nu$, the doping-induced contribution to $\gamma_\nu$ is equal to $\gamma^{\rm (e)}_{\nu\nu\nu\nu}$ in Eq.~(\ref{eq:nonlin_params}).

To find the dependence of the mode frequency on the vibration amplitude we go through the following steps. First, we find the normal modes of interest for the given geometry of the system, with account taken of the boundary conditions, and normalize the displacements $\ub^{(\nu)}(\rb)$  as indicated in Eq.~(\ref{eq:normalization}). We use $\ub^{(\nu)}(\rb)$  to find the strain tensor $\hat\ep^{(\nu)}(\rb)$. The result is substituted into Eq.~(\ref{eq:nonlin_params}) and is convoluted with tensor $\widehat\Lambda_4$, giving the value of $\gamma_\nu$, which is then used in Eq.~(\ref{eq:Duffing_shift}) to find the frequency dependence on the vibration amplitude $\delta\omega_\nu$. 

Of particular interest is the relative frequency shift $\delta\omega_\nu/\omega_\nu$. To find this shift to the leading order, one can disregard nonlinearity when calculating the eigenfrequency $\omega_\nu$.  Then, from Eq.~(\ref{eq:Duffing_shift}),
\begin{align}
\label{eq:scaling}
\frac{\delta\omega_\nu}{\omega_\nu} = \frac{3\gamma_\nu A_\nu^2}{8\int d\rb \widehat\Lambda_2^{\rm (f)} \cdot\hat\ep^{(\nu)}\otimes\hat\ep^{(\nu)}},
\end{align}
where $\widehat\Lambda^{\rm (f)}_2$ is the full tensor of linear elasticity, which includes the major term of the linear elasticity of the undoped crystal and the doping-induced correction $\widehat\Lambda_2$. 

 An important feature of the relative shift $\delta\omega_\nu/\omega_\nu$  is its scaling with the size of the system.
The vibration amplitude $A_\nu$ in Eq.~(\ref{eq:scaling})  can be scaled by the lateral dimension $L$, for example the length of a nanobeam or a nanowire for an extension mode, or the size of the square for a Lam\'e mode, or the diameter of a disk for a breathing mode in a disk. Respectively, we write $A_\nu = \eta_\nu L$.  Then, if one takes into account the explicit form (\ref{eq:nonlin_params}) of the parameter $\gamma_\nu = \gamma^{\rm(e)}_{\nu\nu\nu\nu}$, one finds from Eq.~(\ref{eq:scaling}) that the ratio $\delta\omega_\nu/(\eta_\nu^2\omega_\nu)$ is independent of the system size for the aforementioned modes. In this estimate we used that the tensors $\widehat \Lambda$ are material parameters and are independent of the geometry. We also used that the modes of interest have typical wavelength $\sim L$, and therefore $\hat\ep^{(\nu)}$ scales as $L^{-1}$. 

 Most of the experiments in nano- and micromechanics are done with nanobeams, nanowires, membranes, or thin plates. In such systems the thickness is much smaller than the length or, in the case of membranes or plates, the lateral dimensions. Then, from the boundary condition of the absence of tangential stress on free surfaces \cite{LL_Elasticity}, it follows that the strain tensor $\hat\ep$ weakly depends on the coordinate normal to the surface. This simplifies the denominator in Eq.~(\ref{eq:scaling}), making it proportional to the thickness. Similarly, from Eq.~(\ref{eq:nonlin_params}) $\gamma_\nu$ is also proportional to the thickness, and the thickness drops out of Eq.~(\ref{eq:scaling}).

The explicit expressions for $M\omega_\nu^2$ and $\gamma_\nu$ that determine the denominator and the numerator in Eq.~(\ref{eq:scaling}), respectively, are given in Appendices~\ref{sec:Lame} and \ref{sec:extension} for Lam\'e and extension modes. These expressions are cumbersome, and it is convenient to use symbolic programming to obtain them. \footnote{The program that performs the analytical calculations and evaluates the numerical values of the parameters is available at %\url{http://www.pa.msu.edu/people/dykman/nonlinear_elasticity}.
}

%%%%%%%%%%%%%%%%%%%%%%%%%%%%%%%%%%%%%%%%%%%%%%%%%%%%

\subsection{Temperature and electron density dependence of the scaled nonlinear frequency shift}
\label{sec:T_dependence}

The scaled ratio $\delta\omega_\nu/(\eta_\nu^2\omega_\nu)$ that characterizes the relative nonlinear frequency shift is shown in Fig.~\ref{fig:n_dependence} for several modes that are often used in single-crystal silicon MEMS. This ratio depends on the type of the mode and the crystal orientation. Figure~\ref{fig:n_dependence} refers to high-symmetry crystal orientations, in which case the modes have a comparatively simple spatial structure and the surfaces can be made smooth. We used the values $\Xi_u=8.8$~eV \cite{Yu2001}, $\Xi_{\rm sh} = 300$~eV, the effective mass for density of states $m_{\rm eff} =0.32 m_e$ \cite{hensel1965cyclotron}, and the temperature-dependent linear elasticity parameters given in Ref.~\cite{Varshni1970}.

Figure~\ref{fig:n_dependence} shows that the electron-redistribution induced nonlinearity of vibrational modes is very strong. For the ratio of the vibration amplitude to the system size $\eta \sim 10^{-4}$ and the mode eigenfrequency $\omega_\nu/ 2\pi \sim 10$~MHz, the frequency change can be as a large as $\delta\omega_\nu /2\pi \sim 0.1$~kHz. This explains, qualitatively,  the observations \cite{Yang2016}. A quantitative comparison with the experiment  \cite{Yang2016} is complicated, as the observations refer to different samples. Our preliminary results show an excellent quantitative agreement with the data obtained for the same sample at different temperatures and for different types of modes \cite{Heinz2016}.

The nonlinear frequency shift displays several characteristic features, as seen from Fig.~\ref{fig:n_dependence}. One of them is the strong dependence of the shift on the type of the mode and the crystal orientation. For both the Lam\'e and the extension mode, the shift is much stronger for crystals cut out in
$\langle 100\rangle$ direction than in $\langle 110\rangle$ direction. This is a consequence of the electron energy valleys lying along the $\langle 100\rangle$ axes, making the system more ``responsive" to the lattice displacement along  these axes. Interestingly, in the both configurations the shifts for the Lam\'e  modes are  larger than for the extension modes.
 
A somewhat unexpected feature is the nonmonotonic dependence of the nonlinear frequency shift on the electron density and temperature. The nonmonotoncity occurs in the range where the electron system is close to degeneracy, $\mu_0/k_BT\sim 1$, and it strongly depends on the crystal orientation.  It is much stronger for crystals cut in $\langle 100\rangle$  than $\langle 110\rangle$ directions. For a crystal cut in $\langle 110\rangle$ direction, both the density and temperature dependence of the shift are monotonic in the case of the Lam\'e mode, whereas for the extension mode the nonmonotonicity is weak. 

The nonmonotonicity of the frequency shift stems from the behavior of the parameters $nC_{2,3,4}$ in the range $\mu_0\sim k_BT$.  As seen from Table~\ref{table:tensors}, parameter $nC_2$ exponentially increases with the increasing $\mu_0/k_BT$ for negative $\mu_0/k_BT$, but for large positive  $\mu_0/k_BT$ it falls off as  $(\mu_0/k_BT)^{-3/2}$. It has a pronounced maximum for $\mu_0/k_BT\approx 0.6$. Parameter $nC_3$ also displays a maximum, which occurs for $\mu_0/k_BT \approx 1.1$. In contrast, parameters $nC_{1,4}$ depend on $\mu_0/k_BT$ monotonically. 

The results of Appendices~\ref{sec:Lame} and \ref{sec:extension} show that, for the Lam\'e and extension modes in crystals cut in  $\langle 100\rangle$  direction,  the relative shift  $\delta\omega_\nu/\omega_\nu$ is determined by coefficient $nC_2$,  which explains the nonmonotonicity of the shift. For crystals cut in $\langle 110\rangle$, the shift of the Lam\'e mode is fully determined by coefficient $nC_4$ and is monotonic, whereas for the extension mode the expression for the shift has contributions from $nC_2$, $nC_3$, and $nC_4$ that partly compensate each other, leading to a comparatively small shift all together and its weak nonmonotonicity.

%%%%%%%%%%%%%%%%%%%%%%%%%%%%%%%%%%%%%%%%%%%%%%%%%%%%
%%%%%%%%%%%%%%%%%%%%%%%%%%%%%%%%%%%%%%%%%%%%%%%%%%%%

\section{Conclusions}
\label{sec:conclusions}

The results of this paper show that the electron-phonon coupling strongly affects the nonlinearity of vibrational modes in semiconductor-based nano- and micromechanical systems. The mechanism of the effect is the strain-induced redistribution of the electrons between the valleys of the conduction band. The redistribution results from lifting the degeneracy of the electron energy spectrum by the strain from a vibrational mode. The analysis refers to the range of temperatures where the rate of intervalley scattering strongly exceeds the frequencies of the considered modes. In this case the valley populations follow the strain adiabatically. 

The change of the valley populations is a strongly nonlinear function of the strain tensor. The respective expansion of the free energy in the strain is an expansion in the strain multiplied by the ratio of the electron-phonon coupling energy (in particular, the deformation potential) to the chemical potential of the electron system or the temperature. This ratio is large, $\gtrsim 10^3$. It is this parameter that makes the nonlinearity of the vibrational modes in doped semiconductor structures strong. 

Of special interest in nano- and micromechanical systems is the amplitude dependence of the vibration frequency. To the leading order, it is determined by the quartic terms in the expansion of the free energy in strain. These terms are comparatively large in doped crystals. 

We have calculated the nonlinear elasticity tensor that describes the electron contribution to the terms in the free energy, which are  quartic in the strain. The explicit expressions for the tensor components refer to semiconductors with the valleys on $\langle 100\rangle$ axes, in particular, to silicon. We have also found this tensor for germanium. In  silicon, along with the deformation potential coupling, an important  role is played by the coupling to shear strain. Such strain  lifts the band degeneracy at the zone boundary and is effectively described by a two-phonon coupling. We show that this coupling also leads to strong nonlinearity of vibrational modes. 

The parameter of the electron coupling to shear strain in silicon is not easy to access in the experiment \cite{Hensel1965,Laude1971}. Measurements of the nonlinear frequency shift provide a direct means for determining this parameter. In particular, the nonlinear frequency shift of the fundamental Lam\'e mode in a silicon plate cut along $\langle 110\rangle$ axes is determined by this parameter only, except for small corrections from the nonlinearity of the undoped crystal.

We found that the nonlinear  frequency shift strongly depends on the type of a vibrational mode and the crystal orientation. We also found that the ratio of the frequency shift to the squared vibration amplitude can be profoundly nonmonotonic as a function of electron density and temperature.  The results provide an insight into the experimentally observed strong mode nonlinearity in doped crystals \cite{Yang2016}. In terms of applications, they enable choosing the appropriate range of doping and the temperature regime to optimize the operation of nano- and micromechanical resonators.

%%%%%%%%%%%%%%%%%%%%%%%%%%%%%%%%%%%%%%%%%%%%%%%%%
%%%%%%%%%%%%%%%%%%%%%%%%%%%%%%%%%%%%%%%%%%%%%%%%%

\acknowledgments
We are grateful to T. Kenny for attracting our attention to the problem and for stimulating discussions. We benefited from useful discussions with J. Atalaya, D. Heinz, P. Polunin, S. W. Shaw, and Y. Yang. This research was supported in part by the US Defense Advanced Research Projects Agency (Grant No. FA8650-16-1-7600).

 \appendix
\section{Expansion of the free energy in terms of the strain-induced shift of the energy valleys}
\label{AppendixA}

The major effect of a strain on the electron free energy comes from the shift of the energy valleys. We will assume that  valley $\alpha$ is shifted in energy by $\delta E_\alpha$ and the shift is small, $|\delta E_\alpha|\ll \max(k_BT,\mu_0)$, where $\mu_0$ is the chemical potential in the absence of strain. We further assume that the vibrations are slow compared to the time it takes the electron system to come, locally, to thermal equilibrium for given values of $\delta E_\alpha$, i.e., the temperature and the chemical potential are the same in all valleys. Since for high electron densities the thermal conductivity is high, the change of the temperature compared to the ambient temperature can be disregarded; also, as mentioned in the main text, the electron density $n$ summed over all valleys is constant.

\begin{widetext}
Expanding the electron free energy density to the 4th order in the strain-induced shifts $\delta E_\alpha$, we find that, in an $N$-valley semiconductor, the change $\delta F_e$ of the free energy density is
\begin{align}\label{eq:total_F}
	\frac{\delta F_e}{nk_BT} =& \bar\Delta_\ep + \frac{1}{2}\frac{F'_{1/2}}{F_{1/2}}\left[(\overline\Delta_\ep)^2 - \overline{ \Delta_\ep^2} \right] + \frac{1}{6}\frac{F''_{1/2}}{F_{1/2}}\left[\overline{ \Delta_\ep^3} -3\overline{ \Delta_\ep^2}\;\overline\Delta_\ep +2(\overline\Delta_\ep)^3 \right] \nonumber\\
	&+\frac{1}{8}\frac{{F''_{1/2}}^2}{F_{1/2}F'_{1/2}}\left[(\overline{ \Delta_\ep^2})^2 -2\overline{ \Delta_\ep^2}\,(\overline\Delta_\ep)^2 + (\overline\Delta_\ep)^4 \right]+
	\frac{1}{24}\frac{F'''_{1/2}}{F_{1/2}}\left[
	4\overline{ \Delta_\ep^3}\;\overline\Delta_\ep - \overline{ \Delta_\ep^4} - 
	6\overline{ \Delta_\ep^2}\,(\overline\Delta_\ep)^2 + 3(\overline\Delta_\ep)^4 
	\right].
\end{align}
\end{widetext}
Here, $\overline{\Delta_\ep^m} = N^{-1}\sum_\alpha (\delta E_\alpha/k_BT)^m$. We use the standard notation $F_{1/2}(x)= \int_0^\infty dy \,y^{1/2}/[1+\exp(y-x)]$; primes indicate differentiation over $x$, for example, $F'_{1/2}\equiv dF_{1/2}/dx$. Function $F_{1/2}$ and its derivatives are calculated for $x=\mu_0/k_BT$. 

Equation (\ref{eq:total_F}) immediately gives the tensors $\widehat \Lambda_n$ of the expansion of the free energy increment (\ref{eq:free_energy_expansion}) if one expresses the shift $\delta E_\alpha$ of the valleys in terms of the strain tensor. In the deformation potential approximation the relation between $\delta E_\alpha$ and $\hat\ep$ is given in the main text, see also Eq.~(\ref{eq:general_shift}) below.

In the case of Si crystals, which are often used in micromechanical resonators, an important contribution to $\delta E_\alpha$ comes from the shear-strain induced splitting of the electron energy bands at the zone boundary. Shear strain does not lead to the valley shift in the deformation potential approximation.  The overall shift of valley $\alpha$, to the lowest order in the coupling that causes it (i.e., to the first order in the deformation potential where its contribution is nonzero and to the second order in the band splitting for shear strain) is \cite{Hensel1965}:
\begin{equation}
\label{eq:general_shift}
	\delta E_\alpha = \sum_{ij}\Xi^{(\alpha)}_{ij}\ep_{ij} -  \Xi_{\rm sh} \ep_{\alpha}^2, \quad \Xi_{\rm sh}=\frac{4\Xi_{u'}^2}{\Delta E}.
\end{equation}
Here we use that silicon has six valleys located at the $\langle 100\rangle$ axes, and we chose the coordinate axes $x,y,z$ along $\langle 100\rangle$. Respectively, the valley index $\alpha$ takes on three values that correspond to the $x,y,z$ axes (the valleys lying on the same axis, but in the opposite directions, are equivalent). The strain $\ep_\alpha$, which enters the second term in the right-hand side of Eq.~(\ref{eq:general_shift}), is a component of the strain tensor $\ep_{ij}$ with $i,j$ such that $i,j \neq \alpha$  and $i\neq j$. The parameter $2\Xi_{u'}$  is the interband matrix element of the electron-phonon coupling calculated for the electron conduction bands $\Delta_1$ and $\Delta_{2'}$ at the $X$ point on the boundary of the Brillouin zone, where the bands cross; $\Delta E$ is the energy separation between the bands $\Delta_1$ and $\Delta_{2'}$ at the value of the wave vector $\kb$ that corresponds to the conduction band minimum. Parameter $\Xi_{\rm sh}$ is the effective deformation potential of two-phonon coupling to shear strain. The numerical value of  $\Xi_{\rm sh}$ is not well known. The experimental data give $\Xi_{u'}\approx 7 - 8$~eV \cite{Hensel1965,Laude1971} and the numerical data on the band splitting give $\Delta E\approx 0.7 $~eV \cite{Malone2013} so that $\Xi_{\rm sh}$ is in the range of $280 - 360$~eV; this is essentially an order of magnitude estimate. 

In calculating $\delta F_e$ in Eq.~(\ref{eq:total_F})  we kept terms that are quartic in $\hat\ep$. The components of the tensors $\widehat\Lambda_k$ in Eq.~(\ref{eq:free_energy_expansion}) are expressed in terms of $\delta F_e$ as	
\begin{equation}
	(\Lambda_k)_{i_1j_1...i_kj_k} = \frac{\partial^k \delta F_e}{\partial \ep_{i_1j_1} ... \partial \ep_{i_kj_k}}.
\end{equation}
Tensors $\widehat\Lambda$ are symmetric with respect to the interchange of indices $i_k\leftrightarrow j_k$ and the pairs $(i_kj_k)\leftrightarrow (i_{k'}j_{k'})$. For the considered long-wavelength strain, tensors $\widehat\Lambda_k$ are independent of coordinates. The corrections $\widehat \Lambda_2$ to the linear elasticity tensors were found previously \cite{Hall1967,cerdeira1972effect} and are not discussed in this paper.

%%%%%%%%%%%%%%%%%%%%%%%%%%%%%%%%%%%%%%%%
%%%%%%%%%%%%%%%%%%%%%%%%%%%%%%%%%%%%%%%%%%

\section{Nonlinear elastic constants of germanium}
\label{sec:germanium}

In this section we provide the corrections to the nonlinear elastic constants of germanium, which are due to the redistribution of the electrons  over the valleys. Germanium has four equivalent valleys in the conduction band, which are located on the boundary of the Brillouin zone along $\langle 111\rangle$ axes. We use the Voigt notation and write the components of the corrections to the nonlinear elasticity tensor $\delta\hat c$ in the frame where the axes $(x,y,z)$ are along the $\langle 100\rangle$ directions of the crystal.  Using the results of Appendix~\ref{AppendixA}, we obtain
\begin{align}
\label{eq:Ge}
\delta c_{456} &= \frac{n \Xi_u^{3} F_{1/2}''}{27 F_{1/2} (k_B T)^{2}},\nonumber\\
\delta c_{4444} &= \frac{n\Xi_u^{4}}{81 (k_B T)^3} \left( \frac{3 (F_{1/2}'')^{2}}{F_{1/2} F_{1/2}'} - \frac{  F_{1/2}'''}{F_{1/2}}\right),\nonumber\\
\delta c_{4455} &= \frac{n\Xi_u^{4}}{81 (k_B T)^3} \left( \frac{(F_{1/2}'')^{2}}{F_{1/2} F_{1/2}'} - \frac{  F_{1/2}'''}{F_{1/2}}\right).
\end{align}
The notations are the same as in Appendix~\ref{AppendixA} and in Table~\ref{table:tensors}. The electron-phonon coupling does not contribute to the other third- and fourth-order elastic constants. 

Corrections $\delta c_{44}$ and $\delta c_{456}$ for germanium were found by Keyes~\cite{keyes1961electronic}; however, his final expression for $\delta c_{456}$ differs from Eq.~(\ref{eq:Ge}) by a factor of 4 (our expressions for $\delta c_{44}$ coincide with Ref.~\cite{keyes1961electronic}). Parameters $\delta c_{4444}$ and $\delta c_{4455}$ have not been found before, to the best of our knowledge. In the limiting cases, corrections $\delta c_{4444}$ and $\delta c_{4455}$ have the same dependence on temperature and electron density as constant $nC_2$ discussed in Sec.~\ref{sec:limiting}.

%%%%%%%%%%%%%%%%%%%%%%%%%%%%%%%%%%%%%%%%%%%%%%%%%
%%%%%%%%%%%%%%%%%%%%%%%%%%%%%%%%%%%%%%%%%%%%%%%%%

\section{Duffing nonlinearity parameter for a Lam\'e mode in a square single-crystal plate}
\label{sec:Lame}

We consider a square plate with side $L$ and thickness $h$ made out of a single crystal with cubic symmetry. If the crystal is cut out along $\langle 100\rangle$ or $\langle 110\rangle$ axes,  one of the simplest modes is the first Lam\'e mode \cite{Graff1991}. The normalized displacement field is
\begin{align}
\label{eq:Lame_displacements}
u^{(\nu)}_x &= \sqrt{2} \cos(\pi x/L)\sin(\pi y/L),  \nonumber \\ 
u^{(\nu)}_y &= -\sqrt{2} \sin(\pi x/L)\cos(\pi y/L).
\end{align}
Here, $x$ and $y$ axes are in the lateral plane along the sides of the square, axis $z$ is perpendicular to the plate and $u^{(\nu)}_z=0$. 
Calculating the strain tensor for the displacement (\ref{eq:Lame_displacements}) and substituting the expressions into Eqs.~(\ref{eq:nonlin_params}) and the relation 
\begin{equation}
\label{eq:frequency}
M\omega_\nu^2 =\int d\rb \widehat \Lambda^{\rm (f)}_2 \cdot \hat \ep^{(\nu)}\otimes\hat\ep^{(\nu)},
\end{equation}
for the plate cut out along $\langle  100\rangle$ axes we obtain, in Voigt notation for the elasticity tensors,
\begin{align}
\label{eq:Lame100}
&M\omega_\nu^2 = \pi^{2} h \left(c_{11} - c_{12}\right), \nonumber \\
&\gamma_\nu = \frac{3 \pi^{4} h}{16 L^2} \left( c_{1111} - 4c_{1112} + 3c_{1122}\right).
\end{align}
If we consider silicon and take into account only the contribution $\delta\hat c$ to the nonlinear elasticity tensor $\hat c$,  with the account taken of Table~\ref{table:tensors}, the expression for $\gamma_\nu$ simplifies to 
\begin{align}
\label{eq:Lame100_simplified}
\gamma_\nu = (27\pi^{4} h/32 L^2) \delta c_{1111}.
\end{align}

For the Lam\'e mode cut along the $\langle 110\rangle$ axis, if the tensors are calculated in the axes $\langle 100\rangle$, we have
\begin{align}
&M\omega_\nu^2 = 2 \pi^{2}h c_{44}, \nonumber \\
&\gamma_\nu = (3 \pi^{4}h/2 L^2) \delta c_{4444}.
\end{align}
Note that only  coupling to shear strain contributes to the nonlinearity parameter $\gamma_\nu$ in this case.

%%%%%%%%%%%%%%%%%%%%%%%%%%%%%%%%%%%%%%%%%%%%%%%%%
%%%%%%%%%%%%%%%%%%%%%%%%%%%%%%%%%%%%%%%%%%%%%%%%%

\section{Duffing nonlinearity parameter for an extension mode in a single-crystal  narrow beam}
\label{sec:extension}

We consider the fundamental extension mode in a thin beam of length~$L$ with a rectangular cross-section of area~$S\ll L^2$. The beam is cut along a symmetry axis, and the sides are also along symmetry planes of a cubic crystal. From the free-surface boundary conditions, the normalized displacement field is \cite{Graff1991}:
\begin{align}
\label{eq:displace_extension}
u^{(\nu)}_x &\approx \sqrt{2}\cos(\pi x/L),\nonumber\\
u^{(\nu)}_y &\approx \frac{\sqrt{2} \pi \sigma_2}{L} y \sin(\pi x/L),\nonumber\\
u^{(\nu)}_y &\approx \frac{\sqrt{2} \pi \sigma_3}{L} z \sin(\pi x/L).
\end{align}
This expression takes into account transverse compression that accompanies beam extension and uses the smallness of the beam cross-section; corrections $\sim S/L^2$ are disregarded. The transverse compression in a cubic crystal cut in a symmetric direction is described by Poisson's ratios $\sigma_2$ and $\sigma_3$. Generally, they do not coincide. In Eq.~(\ref{eq:displace_extension}) the transverse coordinates $y$ and $z$ are counted off from the center of the beam 

For the longitudinal direction of the beam $\langle 100\rangle$ and the sides parallel to  $(100)$ planes, the Poisson parameters are equal, $\sigma_2 =\sigma_3$ and $\sigma \equiv \sigma_2=\sigma_3 = c_{12}/(c_{11} + c_{12})$. In this case Eqs.~(\ref{eq:nonlin_params}) and (\ref{eq:frequency}) give
\begin{align}
\label{eq:extension_100} 
M\omega_\nu^2 &= \frac{\pi^{2} S \left(c_{11} \left(c_{11} + c_{12}\right) - 2 c_{12}^{2}\right)}{L \left(c_{11} + c_{12}\right)}, 
\nonumber\\
\gamma_\nu& = (\pi^{4} S/4 L^{3}) \left[ c_{1111} - 8 \sigma c_{1112} \right. \nonumber\\ 
& + 12 \sigma^{2} (c_{1122} + c_{1123}) - 8 \sigma^{3} (c_{1112} +3 c_{1123})  \nonumber \\ 
&\left. + 2 \sigma^{4} (c_{1111} + 4 c_{1112} + 3 c_{1122}) \right].
\end{align}

The expression for $\gamma_\nu$ is simplified if in the nonlinear elasticity tensors we take into account only the contribution  from the electron-phonon coupling as given in Table~\ref{table:tensors} and also allow for the interrelation between different components of the tensor $\delta\hat c$. Then for a silicon beam
\begin{align}
\label{eq:extension_100_simplified}
\gamma_\nu= (\pi^{4} S/4 L^{3})(1+\sigma)^4 \delta c_{1111}.
\end{align}

\begin{widetext}
 For extension along $\langle 110\rangle$ axis, with one side parallel to $(100)$ plane and the other side parallel to $(1\bar 10)$ plane, the Poisson's ratios $\sigma_2 = \sigma(110, 1\bar 1 0)$ and $\sigma_3 = \sigma(110, 001)$ are given in Ref.~\cite{Baughman1998}. 
Then Eqs.~(\ref{eq:nonlin_params}) and (\ref{eq:frequency}) give
\begin{align}
\label{eq:extension_110}
M\omega_\nu^2 &= \frac{4\pi^{2} S}{L} \frac{c_{44} \left(c_{11}(c_{11} + c_{12}) - 2 c_{12}^{2}\right)}{c_{11}(c_{11} + c_{12} + 2 c_{44}) - 2 c_{12}^{2}},\nonumber\\
\gamma_\nu &= \frac{\pi^{4} S}{32 L^3} \Bigl[ c_{1111} \left(\sigma_2^{4} - 4 \sigma_2^{3} + 6 \sigma_2^{2} - 4 \sigma_2 + 8 \sigma_3^{4} + 1\right)\bigr. \nonumber \\
 &+ 4 c_{1112} \left(\sigma_2 - 1\right) \left(\sigma_2^{3} + 2 \sigma_2^{2} \sigma_3 - 3 \sigma_2^{2} - 4 \sigma_2 \sigma_3 + 3 \sigma_2 + 8 \sigma_3^{3} + 2 \sigma_3 - 1\right) \nonumber \\ 
 &+ 3 c_{1122} \left(\sigma_2 - 1\right)^{2} \left(\sigma_2^{2} - 2 \sigma_2 + 8 \sigma_3^{2} + 1\right) + 24  c_{1123} \sigma_3 \left(\sigma_2 - 1\right)^{2} \left(\sigma_2 + \sigma_3 - 1\right) \nonumber \\
 &+ 48 c_{1144} \sigma_3^{2} \left(\sigma_2 + 1\right)^{2}  + 96 c_{1244}\sigma_3 \left(\sigma_2 - 1\right) \left(\sigma_2 + 1\right)^{2} + 24 c_{1155} \left(\sigma_2^2 - 1\right)^{2} \nonumber \\
 &\left. + 24 c_{1266} \left(\sigma_2^2 - 1\right)^{2} + 8 c_{4444} \left(\sigma_2 + 1\right)^{4}\right].
\end{align}
If in the nonlinear elasticity tensor $\hat c$ we take into account only the contribution $\delta\hat c$ from the electron-phonon coupling, in the case of a silicon beam the expression for $\gamma_\nu$ simplifies to 
\begin{align}
\gamma_\nu = \frac{\pi^{4} S}{32 L^3} \left(\left(\sigma_2 - 2 \sigma_3 - 1\right)^4 \delta c_{1111} + 24 \left(\sigma_2 + 1\right)^2 \left(\sigma_2 - 2 \sigma_3 - 1\right)^2\delta c_{1144} + 16 \left(\sigma_2 + 1\right)^4 \delta c_{4444}\right).
\end{align}
Expressions (\ref{eq:extension_100}) and (\ref{eq:extension_110}) were generated using a computer code to calculate the sums and integrals in Eq.~(\ref{eq:nonlin_params}). 
\end{widetext}

%%%%%%%%%%%%%%%%%%%%%%%%%%%%%%%%%%%%%%%%%%%
%%%%%%%%%%%%%%%%%%%%%%%%%%%%%%%%%%%%%%%%%%%

\bibliographystyle{apsrev4-1}
%\bibliography{../../Refs/md10,../refs/Doped-Si/bib/doping_nonlinear}

\end{document}